# THE OPTIMIZATION AND APPLICATION OF THE PROPAGATED RIEMANNIAN WAVEFIELD EXTRAPOLATOR IN VTI MEDIA-PSEUDO-DEPTH DOMAIN LEAST-SQUARES REVERSE-TIME MIGRATION

Hussein Muhammed[1,ID], Sun Xiaodong*[1], Li Zhenchun[1], Abdel Hafiz Gad El Mula[2]

1.China University of Petroleum (East China), School of Geosciences, Department of Geophysics, Qingdao 266580, China.
2.University of Khartoum, Faculty of Science, Department of Geology, P.O. Box 321 Khartoum 11115, Sudan

*Corresponding address: China University of Petroleum (East China), No. 66, W Changjiang Rd., Huangdao District, Qingdao 266580, Shandong, P.R China.（Sun Xiao-Dong: wanliliuyun@163.com / sunxd@upc.edu.cn）

**中文摘要**

LSRTM 的总体框架包括两个步骤；第一个是生成 RTM 图像，第二个是应用最小二乘迁移，但是，这两种操作的收敛会消耗大量时间来提取最终的最小二乘逆时迁移图像，而且在模拟时会产生过采样数据。仅对地震数据应用逆时偏移将根据所应用的成像条件生成带有一些偏移伪影的结果。为了克服这个困境，通过 Born 建模和共轭梯度算法将最小二乘逆时迁移应用于迁移部分。随着速度随深度降低而产生的垂直横向各向同性 (VTI) 介质显着扭曲了逆时偏移结果。这个问题可以通过应用适当的波场外推器在笛卡尔或伪深度域中应用最小二乘逆时偏移来克服。最小二乘逆时偏移外推重建波场使用二维恒定密度声波方程变换到黎曼域，通过均匀采样处理地震信号的过采样效应，并允许在最终偏移图像中恢复更多振幅。对于笛卡尔坐标 (x, y, z) 中的每个点，都有一个对应的垂直时间点，坐标为 $(\xi_1, \xi_2, \xi_3)$，因此我们可以在新的射线坐标中插入重建的源波场通过绘制笛卡尔-黎曼映射函数。不管应用的有限差分格式和具体的边界条件如何，黎曼波场外推器都有一个广义的有限差分公式。有限差分黎曼波场外推器以较低的成本作用于 Born 建模地震数据，产生与经典 LSRTM 精确相似的结果，但由于各种实现问题和后者的过采样效应，出现了一些幅度差异。结果支持域转换策略有效地减少了计算时间而不影响常规 LSRTM 结果的准确性。

**关键词:** 有限差分建模、黎曼波场、伪深度域中的 LSRTM、各向同性和 VTI 介质。

**Abstract**

The general framework of LSRTM consists of two steps; the first one is generating the RTM image and the second is applying the Least-Squares Migration, however, the convergence of both operations consumes a lot of time to extract the final Least-Squares Reverse-Time Migration image and moreover generates oversampling when simulating the data. Applying Reverse-Time Migration to seismic data will generate results with some migration artifacts depending on the applied imaging conditions. To overcome this dilemma, the Least-Squares Reverse-Time Migration is applied to the migrated section through Born modeling and Conjugate Gradient algorithm. Vertical transverse isotropy (VTI) media yielded as the velocity decreases with depth which distorts the Reverse-Time Migration results significantly. This problem can be overcome by applying the Least-Squares Reverse-Time Migration in either the Cartesian or pseudo-depth domains by applying a proper wavefield extrapolator. Extrapolation of Least-Squares Reverse-Time Migration reconstructed wavefield using the 2D constant-density acoustic wave equation transformed into Riemannian domain treats the oversampling effect of seismic signals by making even sampling and allows more amplitude to be recovered in the final migrated image. For every point in the Cartesian coordinate (x, y, z) there is a corresponding vertical-time-point with the coordinates $(\xi_1, \xi_2, \xi_3)$, hence we can interpolate the reconstructed source wavefield in the new ray coordinates by drawing a Cartesian-Riemannian mapping function. Regardless of the applied Finite Difference Scheme and the specific boundaries conditions, the Riemannian wavefield extrapolator has a generalized Finite Difference formula. At a reduced cost, the Finite Difference Riemannian wavefield extrapolator acts on the Born modelled seismic data, producing accurately similar results to the classical LSRTM, yet some amplitude differences are appeared due to various implementation issues and oversampling effect in the latter. The results support that the domain transformation strategy effectively reduces the computational time without affecting the accuracy of the conventional LSRTM results.

# Introduction

The seismic trace contains much information that can be associated with changes in rock properties but can be easily masked by an improper processing algorithm. Solutions to the wave propagation problems by Finite Difference (FD) Methods have received considerable attention in the previous years, in the last decades there has been an increasing interest in the description of physical and chemical processes by means of equations involving fractional derivatives and integrals. The power of these mathematical techniques has a broad potential range of application: fluid dynamics, polymer systems, dynamics of protein molecules, seismic wavefield extrapolation and the diffusion of contaminants in complex geological formations; are recently suggested applications field. The wavefield extrapolation methods are particularly attractive for structurally complex subsurface geometries because of the great difficulties encountered in obtaining analytical solutions. Geometries of particular interest in petroleum exploration are those containing sharp corners that generate diffractions. Accurate and efficient modelling of seismic wavefields that accounts for both attenuation and anisotropy is essential for the further development of seismic data processing methods. It is shown that all wavefield extrapolation methods are based on two equations: the Taylor series and the specific wave equation; hence all errors are inherited from these formulae. The powerful method of separation of variables can also be applied to the acoustic wave equation in the same way as for the heat and usual diffusion equations. We know the size of these errors from the Taylor error formula, in addition, there is a magnification of the errors due to the method itself. To investigate this magnification, we need to look more closely at what the Finite Difference Method is doing. For all geophysical inversions including LSRTM, forward modelling is an essential component (Yao et al., 2018). It provides the means to map the models into the geophysical data. Its mathematical expression also provides the theoretical basis for formulating the inversion algorithms, especially the gradient. Migration aims to determine a model of reflectors; however, wave equations do not include such a model parameter representing the reflectors explicitly. Thus, it is necessary to re-parameterise the wave equation for incorporating this parameter into the forward modelling formulas for tau-domain wavefield migrations.

A revolutionary era of theoretical physics and computational mathematics have developed the field of seismic data processing, due to the advancement in computing and numerical modelling techniques (Wang et al., 2014; Yao et al., 2016), and RTM becomes a mature migration algorithm for industrial application (Sava and Fomel, 2003; Zhang et al., 2007; Xu et al., 2011). However, the step of backward propagation of recorded data in RTM applies the adjoint operator of its forward modelling. Furthermore, the imaging condition widely used for RTM is the zero-lag cross-correlation imaging condition (Claerbout, 1971; Claerbout and Doherty, 1972), instead of a deconvolution imaging condition. This is because the zero-lag cross-correlation imaging condition is unconditionally stable, and more convenient for the implementation (Yao and Jakubowicz, 2016). These two limitations, which are the adjoint operator and the zero-lag cross-correlation imaging condition, have an impact on the RTM seismic imaging, decreasing the resolution and amplitude accuracy of the yielded image. To mitigate the limitations as well as keeping the enhanced features of RTM, LSRTM has been introduced. The Born approximation represents the property of the reflectors with impedance perturbation while the Kirchhoff approximation describes the property with reflectivity, and he forward modelling formulas based on both approximations include the angles of wave propagation (Jaramillo and Bleistein, 1999; Bleistein et al., 2005; Khaniani et al., 2016, Yao and Jakubowicz, 2016). However, it is very complicated and inaccurate to measure the propagation direction of waves by numerically solving wave equations, by FDM. As a result, the Born approximation under the assumption of constant density is a shooting angle independent (Kaplan et al., 2010; Dai et al., 2012; Khalil et al., 2013).

The Finite Difference Method is popularly used for solving geophysical problems like wave equation modelling (e.g., Virieux, 1984, 1986; Robertsson et al., 1994; Carcione 1990; Carcione et al., 2002; Moczo et al., 2007, 2011, 2014; Yao 2013; Khalil et al., 2013; Savioli et al., 2017). LSRTM is achieved by finding an optimal model representing the reflectors to fit the observed data in a least-squares sense. Since the inverse operator is used, LSRTM can generate migration images with higher resolution, fewer artefacts, balanced and preserved amplitude more than RTM (Dai et al., 2012&2013; Yao and Jakubowicz, 2012; Yao; 2013; Zeng et al., 2014; Zhang et al., 2015; Tu and Herrmann, 2015; Wu et al., 2016; Sun et al., 2018). The developed Q-LSRTM method based on the first-order viscoacoustic quasi-differential equations by deriving the Q-compensated forward-propagated operators, Q-compensated adjoint operators and Q-attenuated Born modelling operators (Qu et al. 2021). These traveltime equations are somewhat independent of the vertical velocity in transversely isotropic media (TI) media (Alkhalifah et al.,1997). But truncation of the computational domain to a manageable size requires a non-reflecting boundary condition (Mulder, 2021). Alkhalifah et al. (2001), obtained τ-domain wave equation by applying a coordinate transformation to the conventional eikonal equation and then develop the dynamic part by inverse Fourier transform in space and time to formulate a wave equation in the τ-domain. Because of the high-frequency assumption of the eikonal equation, the resulting wave equation is not accurate in amplitudes. Riemannian wavefield extrapolation generalizes solutions to the Helmholtz equation of the acoustic wave-equation to non-Cartesian coordinate systems, such that extrapolation is not performed strictly in the downward direction (Sava and Fomel, 2001; 2005). The conversion of the vertical axis in the seismic data of the conventional domain from depth to vertical time or pseudo-depth creates a non-orthogonal Riemannian coordinate system (Ma and Alkhalifah, 2013), known as pseudo-depth domain, then wavefield extrapolation for isotropic and anisotropic media in the new coordinate frame can improve efficiency with comparison to Cartesian domain extrapolation results. The aims of this research are to improve the projection efficiency of the LSRTM processed data by using a finite difference scheme projected on a tau-mesh instead of a regular Cartesian one, to minimise calculation time, oversampling of the velocity wavefield, produce better imaging results with clearer structures, higher signal to noise ratio (SNR), higher resolution, more balanced and preserved amplitude, removing migration artefacts and diffracted noises (caused by the surface topography). To test this transformation, we provide results from synthetic data and compare them with their corresponding ones in the Cartesian coordinates. LSRTM in the pseudo-depth domain to one way-wave equation can be achieved (after Sun et al. 2017;2018) by converting the velocity field from the depth domain to the pseudo domain and then calculating the wavefield in the pseudo-depth domain. For a fixed-spread acquisition survey with S shots in total, the computational cost of conventional RTM is approximately $S_\alpha$ ($\alpha$ is the cost of one migration operation). Dai et al. (2012) estimated the speedup of phase-encoding LSRTM relative to conventional RTM by: $speedup \approx \frac{S}{2NI}$, where, $N$ is sub-supergathers, $2NI\alpha$ is computational cost and $I$ is the number of iterations. Assuming that each LSRTM iteration takes double cost of one RTM operation. Thus, the speedup is the optimal result that maybe difficult to achieve for practical applications.

**General Framework of LSRTM**

From Born linearized approximation, the seismic wavefield can be regarded as a combination of a perturbation wavefield and a background wavefield. The scalar wave equation in a time space domain is expressed as in Dai and Schuster, (2009) and Sun et al. (2017). The forward Born modelling can be compactly expressed as a matrix-vector multiplication that linearly relates the reflectivity model to the scattered seismic data through the forward Born modeling operator (Jizhong Yang et al. 2016). A standard migrated image can be computed by applying the adjoint of the forward Born modeling operator to the scattered seismic data (Claerbout, 1992). Least-Squares Migration aims to solve the recent reflectivity model by minimizing the difference between the forward modelled data and the recorded data in a least-squares sense. The least-squares has two mathematical approaches for solving; the conjugate gradient and steepest descent algorithms.

The initial migrated image in this process is a Hessian blurred-version of the sough reflectivity image. If the Hessian operator is an identity matrix, then the migration operator will produce the same result as LSM. In many cases, however, the Hessian operator is not an identity matrix, as its main diagonal elements are nonuniform and differ from unity, and the off-diagonal components are non-zero (Chavent and Plessix, 1999; Nemeth et al., 1999; Plessix and Mulder, 2004; Valenciano et al., 2006; Tang, 2009; Ren et al., 2011 IN Jizhong Yang et al. 2016). The more iteration we do the more we minimize the difference in the misfit function as shown in Figs. 5,6,7,8. As the number of iterations increase the migration results converge to real underground image as shown in the Marmousi model in Figs. 1,2,3,4. In this experiment the 2D constant-density acoustic wave equation using a regular-grid 4th-order space and 2nd-order time FD scheme with Perfectly-Matched Layer (PML) boundary conditions is used to build the RTM data (the program is provided by the second author). We clearly can see that Fig.4 has the perfect results of the Marmousi model, however, this program is time consuming as the results took 20 hours and 23 minutes to reach the optimum Marmousi reflectivity image.

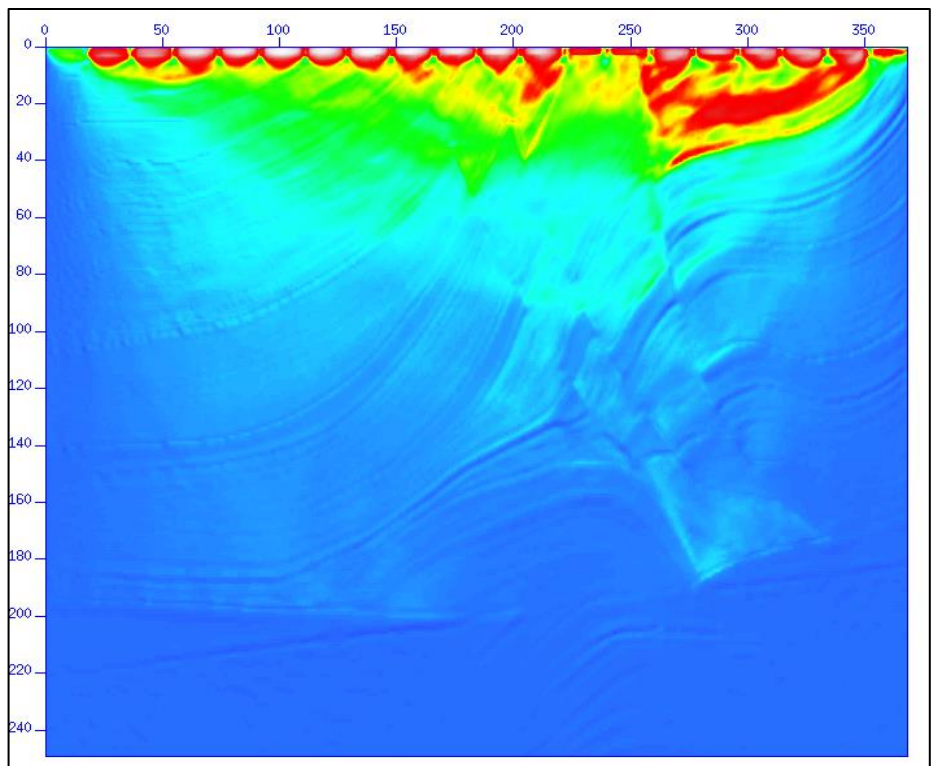

Fig.1: LSRTM Image after **1** iteration, between simulated Marmousi data and initial Hessian operator to find the accurate earth reflectivity image.

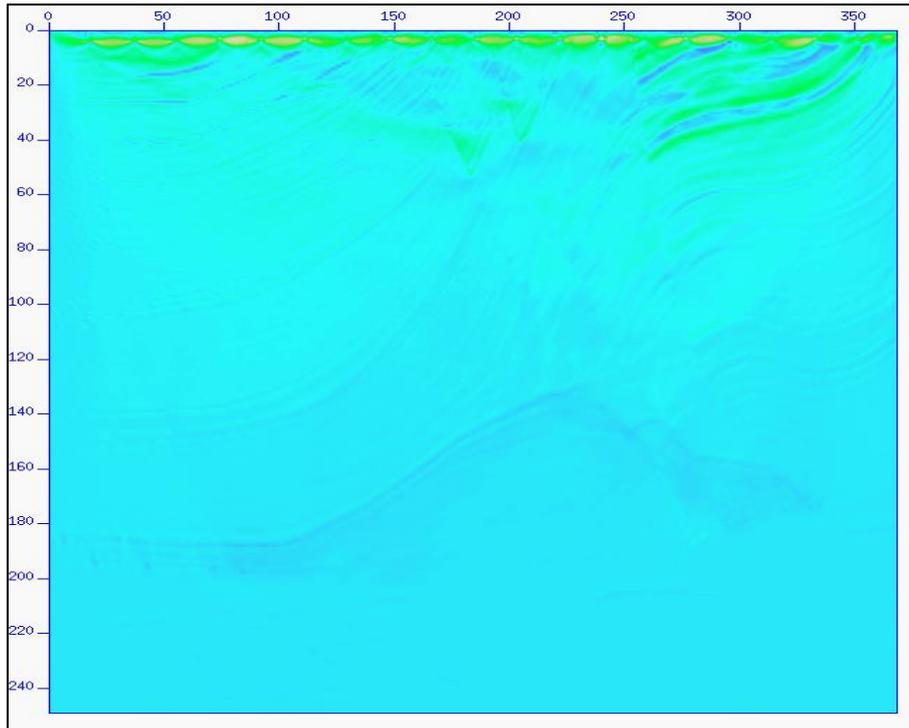

Fig.2: LSRTM Image after **5** iterations, between simulated Marmousi data and initial Hessian operator to find the accurate earth reflectivity image.

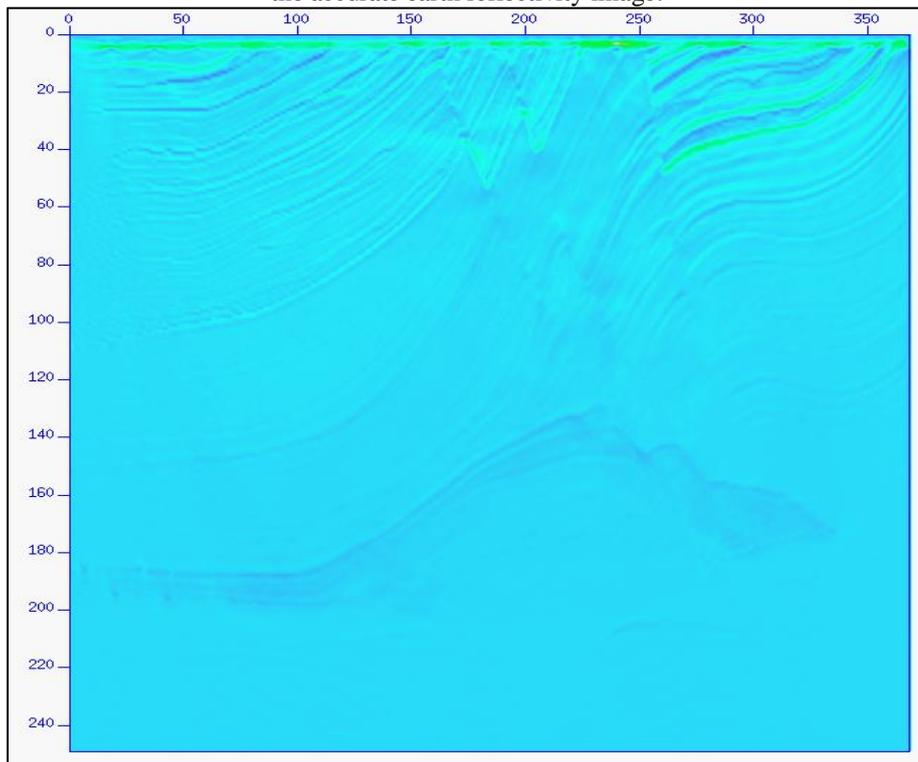

Fig.3: LSRTM Image after **16** iterations, between simulated Marmousi data and initial Hessian operator to find the accurate earth reflectivity image.

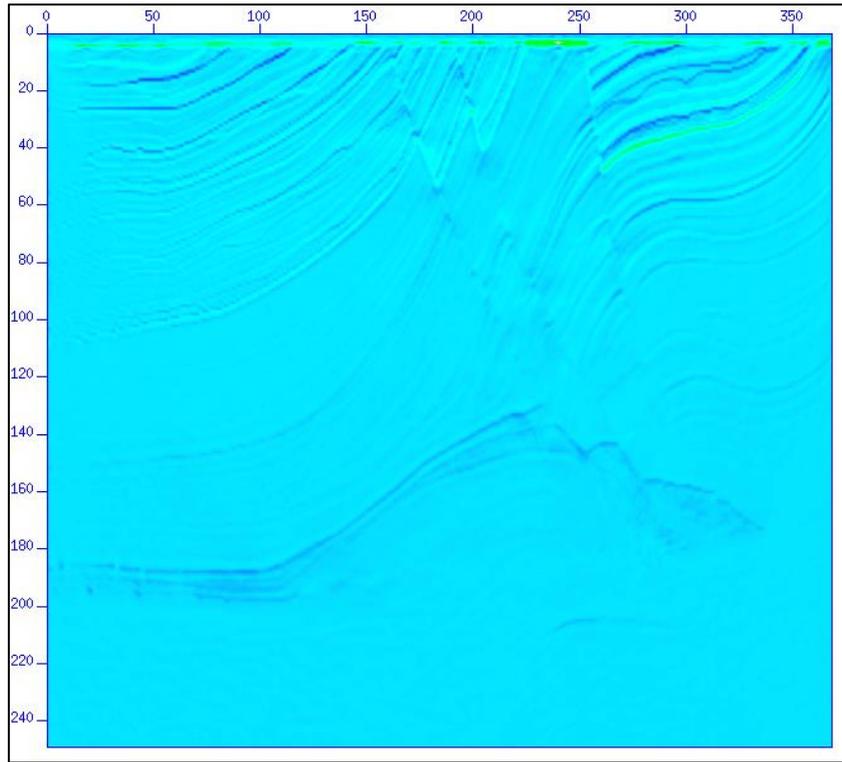

Fig.4: LSRTM Image after **30** iterations, between simulated Marmousi data and initial Hessian operator to find the accurate earth reflectivity image.

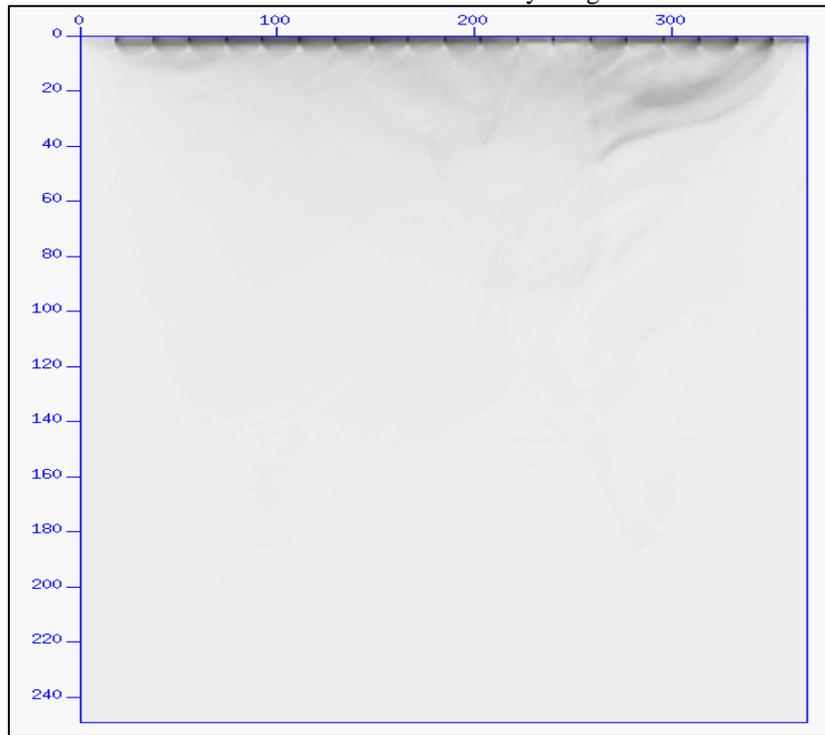

Fig.5: Gradient difference, after **1** iteration, between simulated Marmousi data and initial Hessian operator to find the accurate earth reflectivity image.

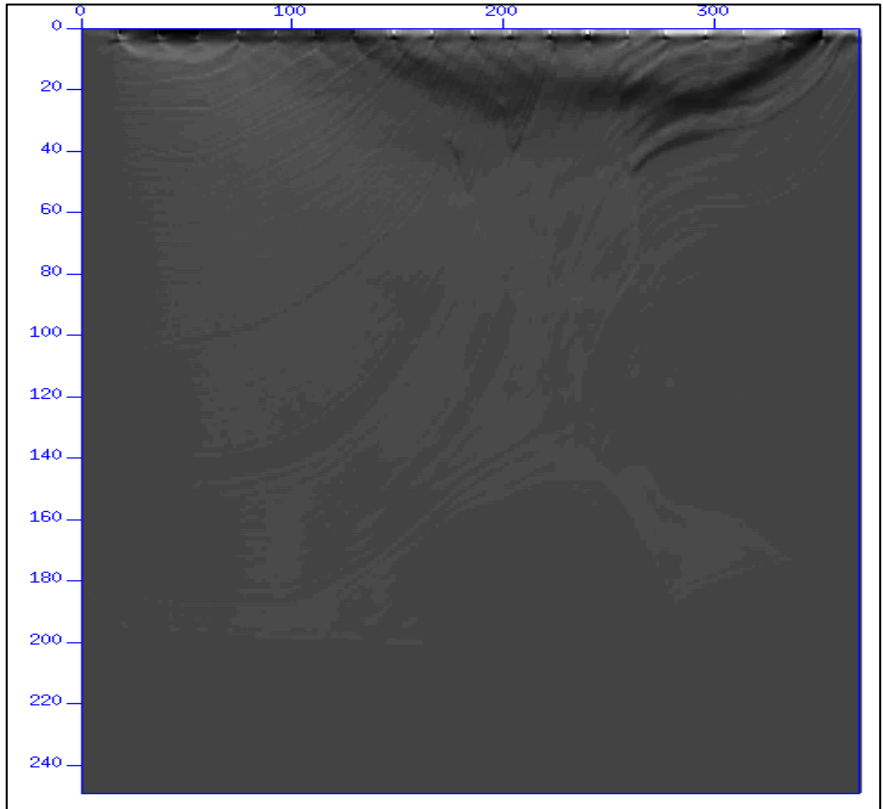

Fig.6: Gradient difference, after **5** iterations, between simulated Marmousi data and initial Hessian operator to find the accurate earth reflectivity image.

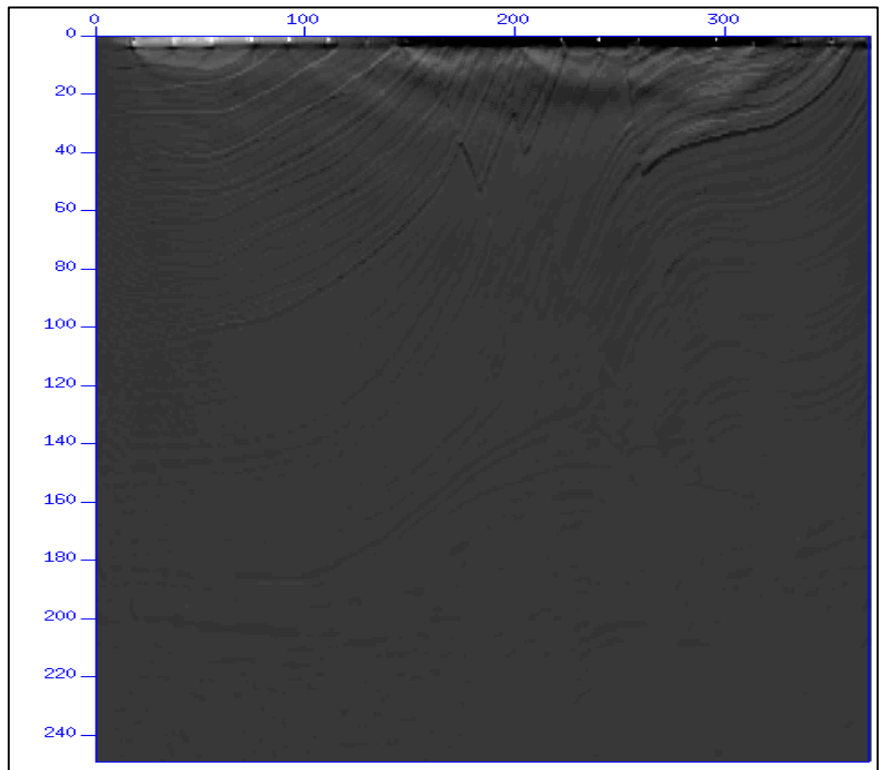

Fig.7: Gradient difference, after **16** iterations, between simulated Marmousi data and initial Hessian operator to find the accurate earth reflectivity image.

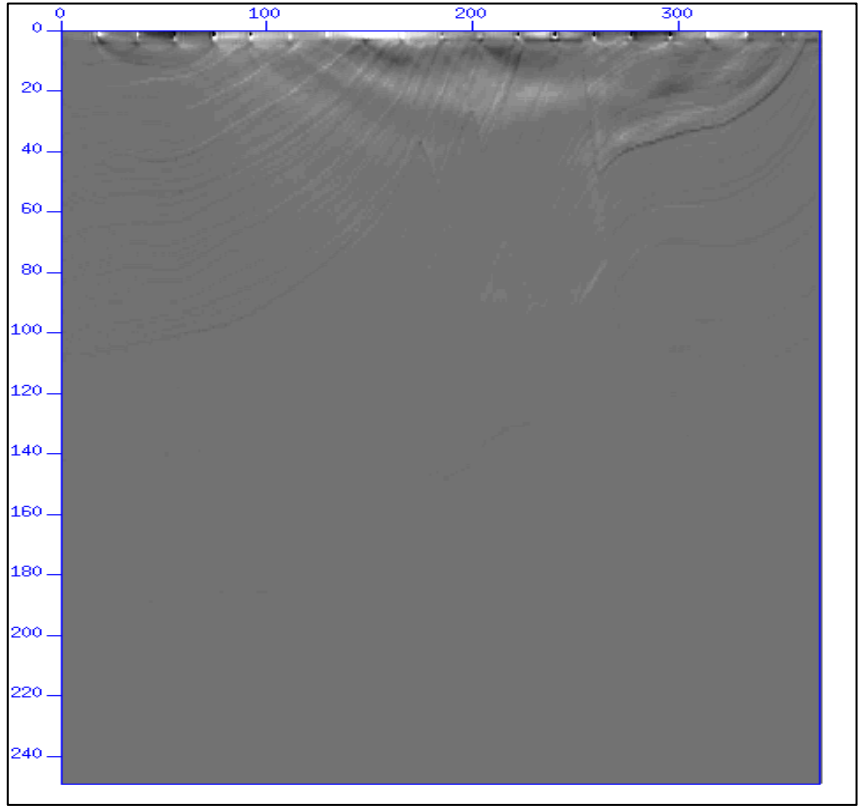

Fig.8: Gradient difference, after **30** iterations, between simulated Marmousi data and initial Hessian operator to find the accurate earth reflectivity image.

## The Reflections-Ray Coordinates System

The conventional seismic section has two axes; one horizontal (distance) and the other is vertical (time). Vertical time axis -tau ($\tau$) is the vertical axis for time-migration (Claerbout, 1985; Yilmaz, 2001; Sava and Fomel 2005; Shragge, 2008; Khalil et al., 2013; Ma and Alkhalifa, 2013; Sun et al. 2018). It is defined as the two-way traveltime measured by coinciding source and receiver on the surface for conventional time image applications, this equation is used in the context of laterally invariant media. To project a depth-point $(x, y, z)$ to vertical-time-point with the coordinates $(\xi_1, \xi_2, \xi_3)$, the vertical axis is converted to time axis $\tau$ via the relationship (Sun et al. (201)8, Ma and Alkhalifa, (2013)):

$$\tau_{TW} = 2 \int_0^z \frac{dz'}{v(z')} \tag{1}$$

The inverse mapping is also straightforward, from differentiation of inverse functions, it follows:

$$z(x, y, \tau) = \int_0^\tau v_m(x, y, \tau') d\tau', \tag{2}$$

the two coordinate systems are related by:

$$\xi_1 = x_1, \quad \xi_2 = x_2, \quad \xi_3 = \tau = \int_0^{x_3} \frac{dx_3'}{v_m(x_1,x_2,x_3')} \tag{3}$$

Based on the above transformation we can interpolate any seismic-space functions between the Cartesian and $\tau$ domains (Ma and Alkhalifa, 2013). The new coordinate system (i.e., $\tau$-domain) is a continuously changing coordinate system and called the ray coordinates (Sava and Fomel, 2005). The new coordinates mapping function can be easily evaluated via ray tracing or Huygens wavefront tracing (Sava and Fomel, 2001; 2005).

## Wave Equation in Riemannian Coordinates

### *i-The Helmholtz Wave Equation*

The Helmholtz wave equation depicts a time-independent form of the wave equation, and resulted from applying the technique of separation of variables, during the application of FD to reduce the complexity of the analysis. The type of the wave equation which is adopted for seismic wavefield simulation is a simplification of the elasticdynamic equations, obtained by putting the shear wave velocity to zero. In the frequency domain on a $\Omega$ domain, the acoustic wave equation can be expressed as:

$$-\frac{\omega^2}{\rho c^2}p - \nabla \cdot \left(\frac{1}{\rho}\nabla p\right) = s, \tag{4}$$

where the pressure, $p(\omega, \mathbf{x})$, depends on the angular frequency $\omega$, the position $\mathbf{x} \in \Omega$, $\rho(\mathbf{x})$ is the density, $c(\mathbf{x})$ is the acoustic speed, and $s(\omega, \mathbf{x})$ represents the source term, $(\nabla)$ is a typical delta function. In the presence of attenuation during propagation of the waves, the complex sound speed also depends on frequency (Aki and Richards, 2002), thus (13) becomes:

$$\frac{1}{c(\omega,\mathbf{x})} = \frac{1}{c_0(\mathbf{x})}\left[1 - \frac{1}{\pi Q}\log\left(\frac{\omega}{2\pi f_{\text{ref}}}\right) + \frac{i}{2Q}\right], \tag{5}$$

where $Q(\mathbf{x})$ is the quality factor and the logarithmic term with reference frequency, $f_{\text{ref}}$,. The complex wavenumber used here in Von-Neumann stability analysis is defined by $k(\omega, \mathbf{x}) = \omega/c(\omega, \mathbf{x})$. Equation (14) supposed to be amalgamated with a suitable absorbing boundary condition, usually consisting in a zero Dirichlet condition. Then, the Helmholtz wave equation for propagating waves in a 3-D Riemannian space becomes (Sava and Fomel, 2004):

$$\frac{1}{\alpha J} = \left[\frac{\partial p}{\partial \zeta}\left(\frac{J}{\alpha}\frac{\partial \mathcal{U}}{\partial \zeta}\right) + \frac{\partial}{\partial \xi}\left(G\frac{\alpha}{J}\frac{\partial \mathcal{U}}{\partial \xi} - F\frac{\alpha}{J}\frac{\partial \mathcal{U}}{\partial \eta}\right) + \frac{\partial}{\partial \eta}\left(E\frac{\alpha}{J}\frac{\partial \mathcal{U}}{\partial \eta} - F\frac{\alpha}{J}\frac{\partial \mathcal{U}}{\partial \xi}\right)\right] = -\frac{\omega^2}{v^2}\mathcal{U} \tag{6}$$

where $\mathcal{U}$ is a scalar function, $\omega$ is temporal frequency, $v[\mathbf{x}(\xi,\eta,\zeta)]$ is the wave propagation velocity, and $\mathcal{U}$ represents a propagating wave, *E, F, G*, and $\alpha$ are differential forms ($J^2 = EG - F^2$). But the wave equation in the new Riemannian coordinates system has two wavefield formulae (Ma and Alkhalifah, 2013) as following:

*ii-Isotropic medium*

A $\tau$-domain two-way wave equation is established by substituting the gradient and divergence operators in the wave equation, resulting the system of equations:

$$\frac{\partial^2 p}{\partial t^2} = v^2 \left(\frac{\partial^2 p}{\partial x^2} + \frac{\partial^2 p}{\partial y^2}\right) + \frac{v^2}{v_m^2}\frac{\partial^2 p}{\partial \tau^2} - \frac{v^2}{v_m^3}\frac{\partial v_m}{\partial \tau}\frac{\partial p}{\partial \tau}. \tag{7}$$

The last term in (7) $\frac{v^2}{v_m^3}\frac{\partial v_m}{\partial \tau}\frac{\partial p}{\partial \tau}$ affects only the extrapolated wavefield's amplitude, thus it can be neglected during the extrapolation operation.

*iii-VTI Medium*

The quasi-P-wave motion in transversely isotropic media with vertical axis of symmetry (VTI) is described by the following second-order system of equations (Duveneck and Bakker, 2011; In Ma and Alkhalifah, 2013):

$$\frac{\partial^2 p_H}{\partial t^2} = (1 + 2\eta)v^2 \left[\left(\frac{\partial}{\partial \xi_1} + \sigma_1 \frac{\partial}{\partial \xi_3}\right)^2 + \left(\frac{\partial}{\partial \xi_2} + \sigma_2 \frac{\partial}{\partial \xi_3}\right)^2\right] P_H + \frac{vv_v}{v_m}\frac{\partial}{\partial \xi_3}\left(\frac{1}{v_m}\frac{\partial p_V}{\partial \xi_3}\right)$$

$$\frac{\partial^2 p_V}{\partial t^2} = vv_v \left[\left(\frac{\partial}{\partial \xi_1} + \sigma_1 \frac{\partial}{\partial \xi_3}\right)^2 + \left(\frac{\partial^2}{\partial \xi_2} + \sigma_2 \frac{\partial^2}{\partial \xi_3}\right)^2\right] P_H + \frac{v_v^2}{v_m}\frac{\partial}{\partial \xi_3}\left(\frac{1}{v_m}\frac{\partial p_V}{\partial \xi_3}\right). \tag{8}$$

where $p_H$ and $p_V$ are horizontal and vertical stresses, respectively, -q is particle momentum, $v_v$ vertical velocity, $v_m$ is mapping velocity, $\eta$ is ellipticity ($\eta = \epsilon - \frac{\delta}{1+2\delta}$), NMO velocity is given by ($v = v_v\sqrt{1 + 2\delta}$) and $\epsilon, \delta$ are Thomsen parameters (Thomsen, 1986; In Ma and Alkhalifah, 2013). We will implement equation (8) in solving the wave propagation in VTI media via FD scheme.

## Boundary Condition and Nonorthogonal FD Solutions

In extracting the scattered seismic wavefield, only a limited model space is simulated, and artificial boundaries are used to truncate the computational model. These artificial boundaries cause pseudo-primaries which in turn produce artefacts within the extrapolated scattered wavefield –in the migrated section. This problem can be solved by using an absorbing boundary condition amalgamated with the Finite Difference scheme. There are many different boundary conditions ready to be use, however, the most accepted and absorbing boundary condition at present is the perfectly-matched layer (PML) (after Berenger 1994; Chew and Liu, 1996; Liu and Tao, 1997; Yao, 2013). The PML condition creates a fictitious material which does not reflect any incident waves and also exponentially decays any waves travelling in it, in addition it removes artificial reflections at the model boundaries, nevertheless it is computationally and memory consuming. The convolutional perfectly matched layer (CPML) boundary condition which is derived for isotropic and anisotropic acoustic wave equations (Pasalic and McGarry, 2010), its equations are expressed as the regular wave equations with additional correction terms. The CMPL boundary condition for VTI media can be expressed as the following system of equations (Pasalic and McGarry, 2010):

$$\frac{\partial^2 P}{\partial t^2} = c_x \frac{\partial^2 P}{\partial x^2} + c_y \frac{\partial^2 P}{\partial y^2} + c_z \frac{\partial^2 P}{\partial z^2} + c_x(\frac{\partial \psi_{p,x}}{\partial x} + \zeta_{p,x}) + c_y(\frac{\partial \psi_{p,y}}{\partial y} + \zeta_{p,y}) + c_z(\frac{\partial \psi_{r,z}}{\partial z} + \zeta_{r,z})$$

$$\frac{\partial^2 R}{\partial t^2} = d_x \frac{\partial^2 P}{\partial x^2} + d_y \frac{\partial^2 P}{\partial y^2} + d_z \frac{\partial^2 P}{\partial z^2} + d_x(\frac{\partial \psi_{p,x}}{\partial x} + \zeta_{p,x}) + d_y(\frac{\partial \psi_{p,y}}{\partial y} + \zeta_{p,y}) + d_z(\frac{\partial \psi_{r,z}}{\partial z} + \zeta_{r,z}). \tag{9}$$

where $P, R$ are pseudo-acoustic wavefields, $c_i$ & $d_i$ ($i \in \{x, y, z\}$) are velocity coefficients, $\psi$ is auxiliary variable with time-evolution and $\zeta_{p,i}$, $\zeta_{p,i}$ are auxiliary variables given by:

$$\zeta_{p,i}^n = a_i \zeta_{p,i}^{n-1} + b_i \left[ \left(\frac{\partial^2 P}{\partial i^2}\right)^n + \left(\frac{\partial \psi_{p,i}}{\partial i}\right)^n \right].$$

the parameters $a_i$ and $b_i$ are given by: $a_i = e^{-(\sigma_i + \alpha_i)\Delta t}$ ; $b_i = \frac{\sigma_i}{\sigma_i + \alpha_i}(a_i - 1).$ (10)

From the above Helmholtz wave equation (Mulder, 2021), a Dirichlet boundary condition for implementing the FD solution is defined on a domain $\Omega$, where; $\Omega = \{(x, z) | x \in [x_{\min}, x_{\max}], z \in [z_{\min}, z_{\max}]\}$, at the rest of the boundary $\Gamma_n = \frac{\partial \Omega}{\Gamma_0}$, we lay a Finite Difference grid to discretize the PDE, with the lowest-order scheme in a 2D case as following:

$$-k_{i,j}^2 p_{i,j} - \frac{p_{i,j}}{\nabla x^2}\left[\frac{p_{i+1,j} - p_{i,j}}{\rho_{i+1/2,j}} - \frac{p_{i,j} - p_{i-1,j}}{\rho_{i+1/2,j}}\right] - \frac{p_{i,j}}{\nabla z^2}\left[\frac{p_{i,j+1} - p_{i,j}}{\rho_{i,j+1/2}} - \frac{p_{i,j} - p_{i,j-1}}{\rho_{i,j-1/2}}\right] = \rho_{i,j} s_{i,j}, \tag{11}$$

The final derivation of the above formula called a numerically exact non-reflecting boundary condition (Mulder, 2021) on one side of the domain. At the end, the FD solution of the equation (7, 8) in a generalized Riemannian coordinates system through Domain-limited FD scheme is given in Khalil et al., (2013) as following:

$$u(i, j, k + 1) = 2u(i, j, t) - u(i, j, k - 1) + c_{nn}(i, j)u(i - 1, j - 1, k + \lambda_\tau) + c_{no}(i, j)u(i - 1, j, k + \lambda_\tau) + c_{np}(i, j)u(i - 1, j + 1, k + \lambda_\tau) + c_{on}(i, j)u(i, j - 1, k + \lambda_\phi) + c_{oo}^\tau(i, j)u(i, j, k + \lambda_\tau) + c_{oo}^\phi(i, j)u(i, j, k + \lambda_\tau) + c_{op}(i, j)u(i, j + 1, k + \lambda_\phi) + c_{pn}(i, j)u(i + 1, j - 1, k + \lambda_\tau) + c_{po}(i, j)u(i + 1, j, k + \lambda_\tau) + c_{po}(i, j)u(i + 1, j + 1, k + \lambda_\tau). \tag{12}$$

The Riemannian-mesh stencil's coefficients are: $c_{nn} = \gamma \frac{\beta_{12}}{2\Delta\tau\Delta\phi}$, $c_{oo}^{\phi} = 2\gamma(\frac{\beta_{22}}{\Delta\phi^2})$, $c_{no} = \gamma(\frac{\beta_{11}}{\Delta\tau^2} - \frac{\alpha_{11}+\alpha_{21}}{2\Delta\tau})$, $c_{op} = \gamma(\frac{\beta_{22}}{\Delta\phi^2} - \frac{\alpha_{22}+\alpha_{21}}{2\Delta\phi})$, $c_{np} = -c_{nn}$, $c_{pn} = -c_{nn}$, $c_{on} = \gamma\left(\frac{\beta_{22}}{\Delta\phi^2} - \frac{\alpha_{22}+\alpha_{12}}{2\Delta\phi}\right)$, $c_{po} = \gamma\left(\frac{\beta_{11}}{\Delta\tau^2} + \frac{\alpha_{11}+\alpha_{21}}{2\Delta\tau}\right)$, $c_{oo}^{\tau} = 2\gamma(\frac{\beta_{11}}{\Delta\tau^2})$, $c_{pp} = c_{nn}$, $\gamma = (\Delta t v)^2$.

The stencil's coefficients are functions of vertical traveltime ($\tau$) and shooting angle ($\phi$) (Sava and Fomel, 2001; 2005), then we can define the coordinate mapping functions as in equation (3), we can estimate the CFL condition for the stability of the scheme roughly by: $CFL = \frac{\sqrt{\gamma}}{\tau.\Delta\phi}$. The extrapolation process can only be limited the reconstructed source wavefield (Khalil et al., 2013) which is sufficient to perform the Pseudo-domain LSRTM.

## Numerical Experiments and Results

Fig.9 represents an impulse response using a homogeneous anisotropic velocity model, generated by constructing a specific computational code recipe, and has a vertical velocity of 2.1 km.s$^{-1}$, NMO velocity of 1.642 km.s$^{-1}$ and an-ellipticity eta = 0.1. Moreover, to the reduced computational cost, tau-domain anisotropic extrapolation also attenuates Cartesian-domain-related-extrapolated shear wave artifacts. Fig.9b shows the removal of the shear wave artefact in tau-domain, with coarser vertical sampling which in turn enhance the numerical dispersion of shear waves, that generated in the Cartesian domain (Fig.9a), while Fig.9c represents the signal differences between the two projection schemes.

Figs.10,11&12 display part of anisotropic Marmousi (VTI) dataset with Pre-stack LSRTM in both Cartesian and Riemannian domains and after conversion back from tau domain, respectively. The original Marmousi data set contains 240 shots with 96 traces per shot and total recording time of 2.9 seconds, this anisotropic dataset has peak frequency of 30 Hz, spread length up to 3.0 km. Parts of the model are potentially invertible below the surface location of 3 km. The Cartesian-based LSRTM in Figs.10&12 obtained by applying Born approximation and CG-algorithm, while the conversion of the RTM reflectivity model to tau domain in Fig.11 treats the model evenly but some still some distortions appear. The final LSRTM result in Fig.12 shows slightly-enhanced image after converting it back to Cartesian domain because of balancing the amplitude and removing artefacts. The results took 5 hours and 30 minutes (measured by a stopping watch) to be obtained on a high-speed PC cluster.

The reservoir appears all its sections between 2.0 km to 3.0 km, but its position is slightly shifted in figure 11, this can be attributed to anisotropic extrapolation of the wavefield on tau domain.

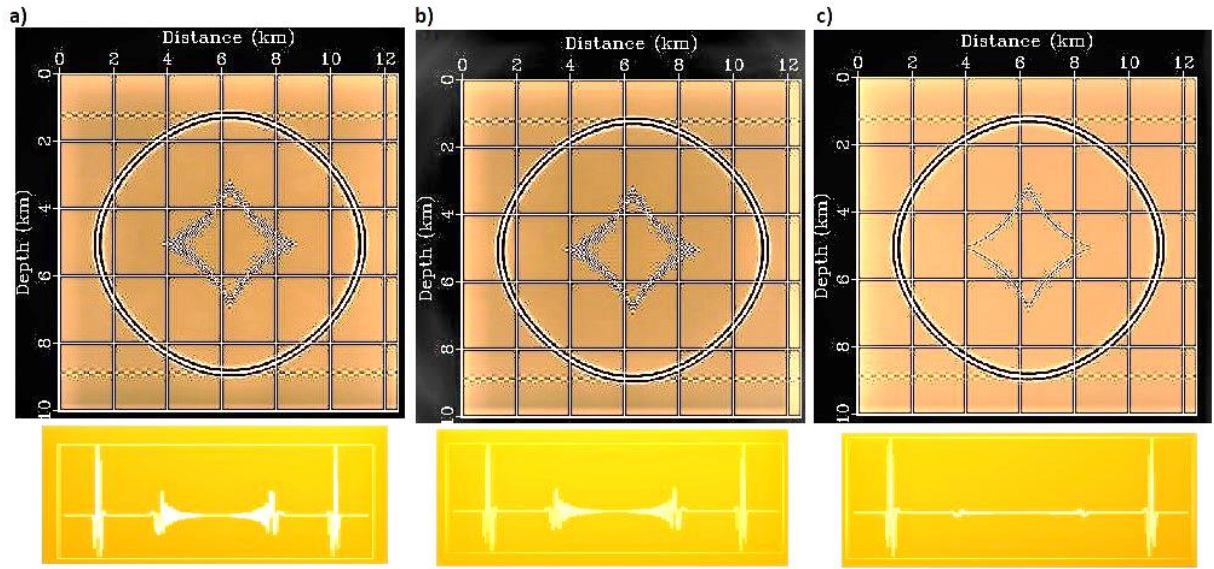

Fig.9: Anisotropic impulse responses and its corresponding signal obtained in: (a) Cartesian and (b) Riemannian domains. (c) The removed shear waves signal after projection on tau mesh.

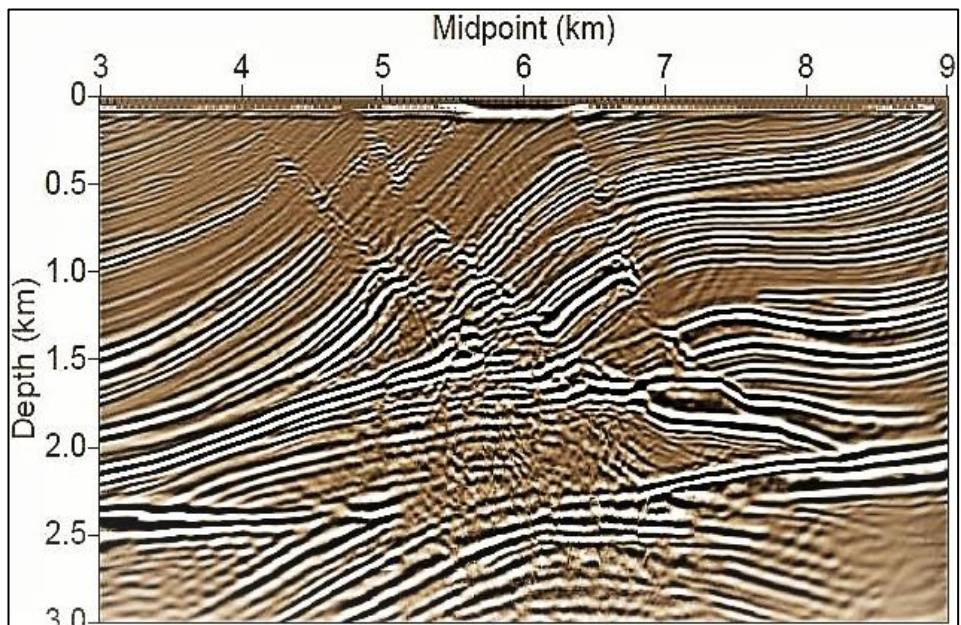

Fig.10: Pre-stack LSRTM for part of anisotropic Marmousi dataset in Cartesian domain.

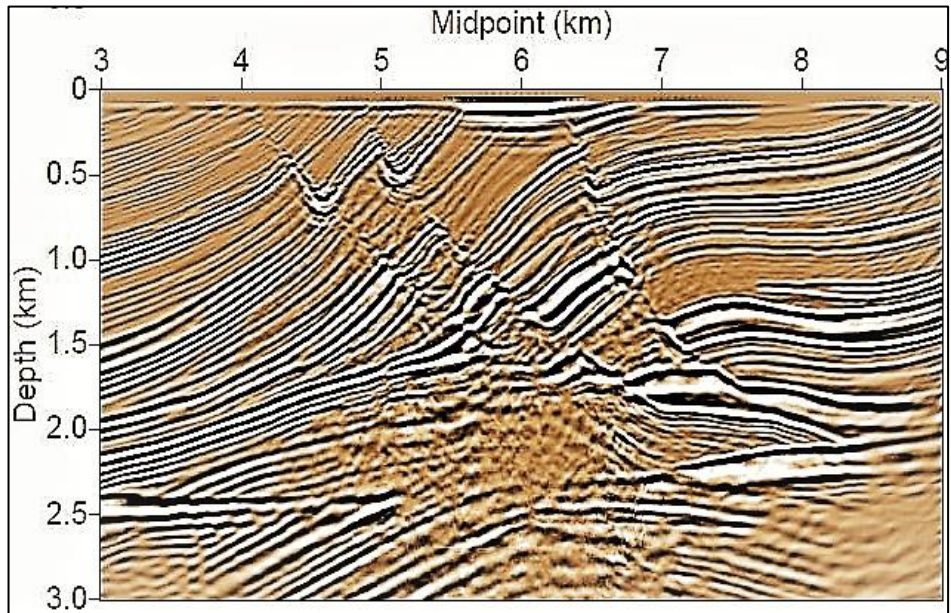
Fig.11: Pre-stack LSRTM for part of anisotropic Marmousi dataset in Riemannian domain.

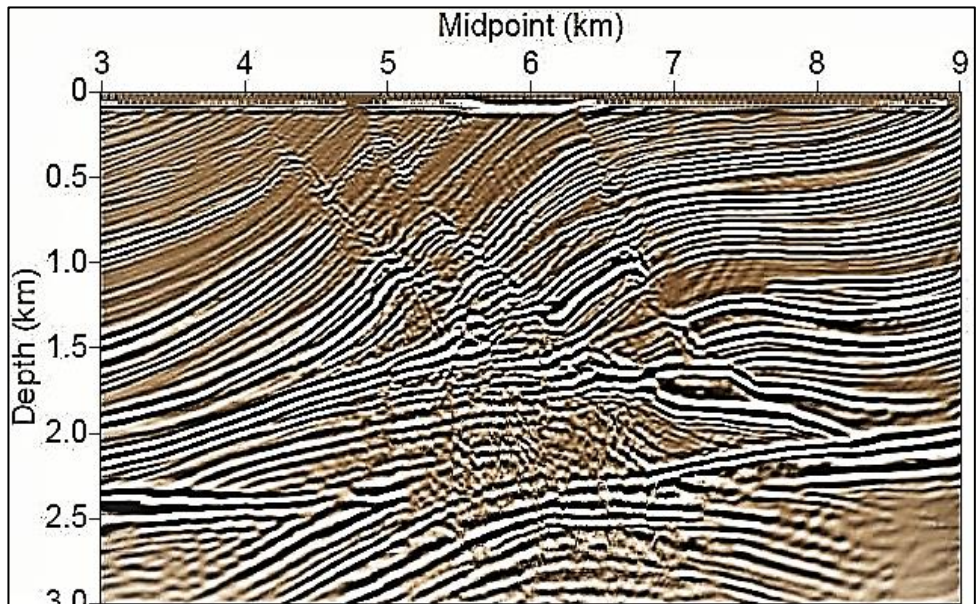
Fig.12: Pre-stack LSRTM for part of anisotropic Marmousi dataset in Cartesian domain after converted back from tau domain.

## Conclusion

Conventional LSRTM gives proper migration results as shown in Fig. 4. Riemannian space by definition is described as a non-orthogonal curvilinear coordinates system. Using it to perform LSRTM, speedup the extrapolation of the anisotropic LSRTM wavefield more than doing it with Cartesian mesh. Despite are the migration results obtained with the same number of iterations, the accuracy of the final migration results almost identical before or after the conversion. However, shear wave artefacts can be removed properly when projecting the VTI seismic data on Riemannian domain. It is therefore evident that further research is required to obtain advanced results, including processing 3D seismic data, stability analysis of the FD scheme, viscoacoustic LSRTM and non-linear LSRTM.

## ACKNOWLEDGMENT

This research gets funded by National Natural Science Foundation of China (41630964), the Fundamental Research Funds for the Central Universities (18CX02059A), the Development Fund of Key Laboratory of Deep Oil & Gas (20CX02111A), Shandong Natural Science Foundation of China (ZR2020MD048) and the Major Scientific and Technological Projects of CNPC under grant ZD2019-183-003.

**Hussein Abduelhaleim Hussein Muhammed, (18 April 1993), male, telephone: +86-15621423175**

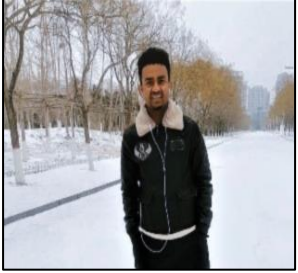

B.Sc. (Hons.) degree in Geology from the University of Khartoum, Sudan (Oct. 2015). Currently is a master degree student in the Department of Geophysics at China University of Petroleum (East China). His research interests include; Seismic Exploration Methods, FDM & FEM modeling, Seismic waves propagation, imaging and inversion.

**Dr. Xiao-Dong Sun (Lecturer)**

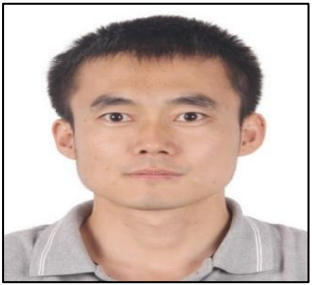

Master's supervisor. Lecturer in Geophysics. Expertise Areas: Seismic propagation, Seismic imaging, Seismic inversion. His research interests: seismic waves; propagation, imaging, inversion, RTM, LSRTM, FWI and inversion methods.

**Professor Dr. Li Zhen-Chun**

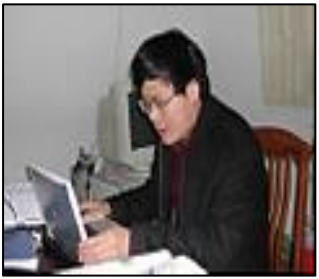

Director of the department of Geophysics, School of Geosciences, China University of Petroleum. Master's and PhD's supervisor. He is the leader of the Seismic Waves Propagation and Imaging (SWPI) research group.

青岛校区： 青岛市黄岛区长江西路 66 号
邮编：266580　网址：http://www.upc.edu.cn/index.htm

# 2021 年"油气•智能 @未来科学家"研究生国际学术论坛（2 号通知）

　　石油与天然气能源在国民经济和社会发展中起着重要作用，人工智能及大数据的发展为深层-超深层和非常规油气勘探开发提供了重要支撑，对复杂领域油气资源进行智能、高效勘探开发已成为重要研究热点。为了进一步拓宽研究生的学术视野，提高研究生的创新能力，搭建国际学术交流平台，中国石油大学（华东）将承办山东省 2021 年"未来科学家"研究生国际学术论坛，拟于 2021 年 11 月上旬在中国石油大学(华东)青岛校区举办。

　　本次"未来科学家"研究生国际学术论坛将围绕"油气+智能"和"人才培养"主题，充分利用中国石油大学（华东）地质资源与地质工程一流学科资源优势和现有的 2 个 创新引智基地等优质资源，邀请国内外知名专家作专题报告，帮助研究生了解学科前沿动态，学习借鉴国内外最新成果，开拓研究生国际视野；搭建研究生国际学术交流平台，促进研究生和国内外知名专家学者的学术交流，通过共同研讨，为提升研究生培养质量贡献力量。

　　本次论坛将设置"油气+智能与未来"主题论坛和"油气勘探"专题论坛，并邀请国际著名学者和刊物主编就学术提升与论文写作进行专题讲座。

　　主题论坛：主要由特邀专家进行主题报告，针对油气能源、人工智能和未来科学发展、技术需求、人才培养等，进行前沿系列讲座。

　　专题研讨：针对深层油气、非常规油气、其他能源矿产等相关的勘探理论方法和技术进行学术交流。

　　论文写作：开放讲习班，分享在高水平国际期刊上发表学术论文的要求和心得，同时推动研究生更好地了解学科国际前沿，开展专业学习和科研工作。

　　实践考察：参观深层油气重点实验室和山东省油气藏重点实验室，进行黄岛海岸地质实习考察。

　　欢迎国内外优秀博士、硕士研究生积极参与。

## 一、会议主题

　　油气•智能 @未来科学家

**1 / 4**

## 二、会议组织

主办单位：山东省教育厅

承办单位：中国石油大学（华东）

## 三、会议内容

1、油气+智能专家特邀报告；

2、油气地质与地球物理学术研讨；

3、论文写作研讨；

4、实践考察。

## 四、会议组织结构

**【学术委员会】**

主　任：郝　芳

副主任：操应长

成　员：唐晓明　刘可禹　蒋有录　符力耘　林承焰　刘　华　胡钦红　印兴耀

　　　　卢双舫　范宜仁　韩同城　宗兆云　黄建平　王　民　孟庆峰

**【指导委员会】**

主　任：刘华东

副主任：阎子峰

成　员：王备战　刘　华　裴仰文　俞继仙　蔡宝平　邓少贵　宗兆云　张广智

　　　　吴智平　郑珊珊　曹丹平　李福来

**【组织委员会】**

主　任：刘　华

副主任：邓少贵　宗兆云　郑珊珊　张　锋

成　员：邱隆伟　张立强　李振春　王　民　谭宝海　王力禾　李　勇　邹桂红

杨建敏　栗　磊　姜永昶

秘　书：李　勇　邹桂红

## 五、会议安排

1、会议时间：2021年11月2-5日（11月2日为报到时间）
2、会议地点：中国石油大学（华东）唐岛湾校区
3、会议交流语言：英语

## 六、会议注册、住宿安排及交通指南

1、会议注册费：本次论坛免注册费。

2、食宿安排：本次会议凡投稿并被录用安排论坛交流人员，食宿由会务组统一安排，双人间标准，其中市内参会人员不提供住宿，参会其它费用请自行承担。

3、报名方式：本次论坛仅设立现场交流发言，不设墙报展示，请参会代表于2021年10月20日前将论文英文全文和会议PPT发送至联系人邮箱 liyong@upc.edu.cn。在截止日期前没有收到PPT的将不再安排现场交流。

4、优秀论文奖：本次论坛将设立优秀论文奖，请需要申报的人员务必同时发送全英文论文至相应邮箱，否则视为放弃本次优秀论文评选。

5、疫情防控要求：根据疫情防控要求，请参会人员提前14天做好健康监测,并按照要求提供绿色健康码和14天行程单并按照要求进行核酸检测后才能进入校园参会。具体防疫要求将在3号通知内进行发布。

6、联系人：

  李　勇　0532-86981750　　liyong@upc.edu.cn
  邹桂红　0532-86981750　　330556513@qq.com
  谭宝海　18615328812　　　tanbaohai@upc.edu.cn
  会议QQ交流群号：137467395，加群请注明学校、姓名

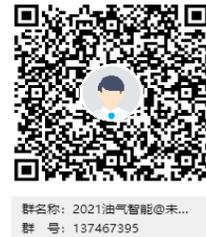

附件 1

## 2021 年"油气·智能 @未来科学家"研究生国际学术论坛回执

| 单位 | | | |
|---|---|---|---|
| 参会人员姓名 | | 参会人员身份 | □博士、□硕士 |
| 通信地址 | | | |
| 联系电话 | | Email | |
| 提交会议报告题目（中英文） | | | |
| 对会议安排与议题的建议 | | | |

此回执复印有效

# 2021 International Academic Forum for Graduate Students of "Oil and Gas • Intelligence @Future scientist"

Oil and gas energies have the important roles in the national economic and social development. Artificial Intelligence and big data analysis are two new and necessary technologies to boost future exploration and development of the deep-superdeep, unconventional oil and gas reservoirs. How to explore and develop the complex oil and gas reservoirs with higher efficiency and intelligence is becoming more and more attention in the industry and scientific area. In order to improve the academic research and innovation ability of graduate students, and establish an international academic exchange platform, the International Academic Forum for graduate students of "Future Scientists" in Shandong will be hosted by China University of Petroleum (East China) in the earlier of November, 2021 at Qingdao campus.

This International Academic Forum of "Future Scientist" for graduate students will focus on the theme of "Oil & Gas, Intelligence @Future Scientists", and making full use of the advance discipline resources of Geological resources and geological engineering of China University of Petroleum (East China) and the two Overseas Expertise platforms, including "Deep-Superdeep Oil & Gas Geophysical Exploration" and "Tight Oil & Gas Geology and Exploration".

Famous Scientists from domestic and overseas will be invited to give special reports, to introduce the frontier and new progress of the discipline, to extend the international perspective on the latest achievements in the world, to build an international academic exchange platform for joint discussions among graduate students. This Form will promote the academic exchanges between graduate students and well-known experts from domestic and overseas and enhance the scientific ability of graduate students.

There are two topics in this forum: "Oil and Gas + Intelligence and Future" and "Oil and Gas Exploration". Renowned scholars and journal editors will be invited to give lectures on academic research and paper writing.

Theme Forum: A series of high-level lectures will be conducted by invited experts on oil and gas, Artificial intelligence, scientific development in the future, technical requirements, and graduates' education.

Special seminar: Academic exchanges will aim to exploration theoretical methods and technologies related to deep oil and gas, unconventional oil and gas, and other energy minerals.

Paper writing: Study groups are organized to share the requirements and experiences on publishing academic papers in the high-level international journals. At the same time, it will help graduate students to learn new knowledge of disciplines in the world and carry out professional study and research work.

Visits and Geological field trip: Visit the key laboratories of the deep oil and gas, the key laboratories of oil and gas reservoir of Shandong Province, and have the geological field trip around Huangdao seaside.

Welcome excellent domestic and overseas Doctor and postgraduate students to participate this Academic Forum.

# 1. Conference Theme
"Oil and Gas • Intelligence @Future scientist"

# 2．Organization:
Sponsor: Department of Education, Shandong Province
Organizer: China University of Petroleum (East China)

# 3. Conference Content
1. Expert lectures on Artificial Intelligence, Oil and gas energy;
2. Academic discussion about the oil and gas geology and geophysics;
3. Paper writing discussion;
4. Visiting and field trip

During the conference, famous experts from domestic and foreign will be invited to give relevant academic reports.

# 4. Conference Organization
**Academic Committee**

Chairman：Hao Fang

Vice Chairman：Cao Yinchang

Member：Tang Xiaoming, Liu Keyu, Jiang Youlu, Fu Liyun, Lin Chengyan, Liu Hua,
　　　　　Hu Qinhong, Yin Xingyao, Lu Shuangfang, Fan Yiren, Han Tongcheng,
　　　　　Zong Zhaoyun, Huang Jianping, Wang Min, Meng Qingfeng
**Steering Committee**

Chairman：Liu Huadong

Vice Chairman：Yan Zifeng

Member：Wang Beizhan, Liu Hua, Pei Yangwen, Yu Jixian, Cai Baoping, Deng Shaogui, Zong Zhaoyun, Zhang Guangzhi, Wu Zhiping, Zhen Shanshan, Cao Danping, Li Fulai

**Organizing Committee**

Chairman：Liu Hua

Vice chairman：Deng Shaogui, Zong Zhaoyun, Zhen Shanshan, Zhang Feng

Member：Qiu Longwei, Zhang Liqiang, Li Zhenchun, Wang Min, Tan Baohai, Wang Lihe, Li Yong, Zou Guihong, Yang Jianmin, Li Lei, Jiang Yongchang

Conference Secretary: Li Yong, Zou Guihong

## 5. Conference Arrangement

1. Conference time: Nov 2$^{nd}$-5$^{th}$, 2021( Nov 2$^{nd}$ , Registration time)
2. Venue: China University of Petroleum (East China), Qingdao, Shandong
3. Conference language: English

## 6. Conference registration and accommodation

1. Registration fee: It's free to attend this forum.
2. Accommodation: The conference will provide the accommodation for invited participants from other cities. The conference will provide free lunch for all participants.
3. Registration and the full paper: Participants shall submit full English paper and presentation PPT to liyong@upc.edu.cn before Oct. 20$^{th}$ 2021.
4. Excellent paper award: This seminar will issue the excellent paper awards from the authors who have submitted the full papers. Authors, who didn't submit the full English paper on time, will be considered to give up this opportunity.
5. COVID-19 prevention requirements: According the regulation of COVID-19 prevention, all attendee shall be have the 14 days health check and take the health code and trip record for examination to attend this meeting. The details will be published later.
6. Conference Contact

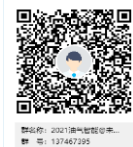

| | | |
|---|---|---|
| Li Yong | 0532-86981750 | liyong@upc.edu.cn |
| Zou Guihong | 0532-86981750 | 330556513@qq.com |
| Tan Baohai | 18615328812 | tanbaohai@upc.edu.cn |

Conference QQ group: 137467395，pls indicate your name and your University

<div align="right">

**China University of Petroleum (East China)**
**Oct 9$^{th}$, 2021**

</div>

# Registration form of 2021 International Academic Forum for Graduate Students of "Oil and Gas • Intelligence @Future scientist"

| University | |  | |
|---|---|---|---|
| Participant Name | | Student type | ☐Doctor ☐Master |
| Address | | | |
| Telephone | | Email | |
| Article title | | | |
| Suggestions | | | |

The copy of this form can also be submitted.

# OGIFS

Oil and Gas · Intelligence @Future Scientist

## THE BEST PAPER AWARD IS PRESENTED TO

## Hussein Muhammed

### For the paper entitled

The optimization and application of the propagated Riemannian wavefield extrapolator in VTI media-pseudo-depth domain least-squares reverse-time migration

2021 International Academic Forum for Graduate Students of
"Oil and Gas • Intelligence @Future Scientist"

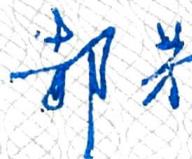

President of China University of Petroleum
Academician of Chinese Academy of Sciences
November 4 2021